\documentclass[12pt]{article}
\usepackage{graphicx}
\usepackage{amsmath}
\usepackage{amssymb}
\usepackage[left=1in,top=1in,right=1in,bottom=1in]{geometry}
\usepackage{natbib}
\usepackage{setspace}
\makeatletter
\renewcommand\section{\@startsection{section}{1}{\z@}{-3.25ex plus -1ex minus -.2ex}{1.5ex plus .2ex}{\normalsize\bf}}
\renewcommand\subsection{\@startsection{subsection}{2}{\z@}{-3.25ex plus -1ex minus -.2ex}{1.5ex plus .2ex}{\normalsize\bf}}
\renewcommand\subsubsection{\@startsection{subsubsection}{3}{\z@}{-3.25ex plus -1ex minus -.2ex}{1.5ex plus .2ex}{\normalsize\bf}}
\makeatother

\newtheorem{thm}{Theorem}

\newtheorem{prop}[thm]{Proposition}

\newcommand{\supp}[1]{\text{supp}(#1)}

\begin{document}
\begin{center}
\textbf{A Brief Remark on Energy Conditions and the Geroch-Jang Theorem}\footnote{I am grateful to David Malament for helpful comments on a previous draft of this paper, including several suggestions both on how to formulate the statements of the propositions in the paper and on how to simplify their proofs. I am also grateful to Bob Geroch and Wayne Myrvold for helpful conversations on topics closely related to this paper.}\\[1\baselineskip]
James Owen Weatherall\footnote{weatherj@uci.edu} \\Logic and Philosophy of Science \\
University of California, Irvine\\[1\baselineskip]
\end{center}
\singlespacing
\begin{center}\textbf{Abstract}\end{center}
The status of the geodesic principle in General Relativity has been a topic of some interest in the recent literature on the foundations of spacetime theories. Part of this discussion has focused on the role that a certain energy condition plays in the proof of a theorem due to Bob Geroch and Pong-Soo Jang [``Motion of a Body in General Relativity.'' \emph{Journal of Mathematical Physics} \textbf{16}(1), (1975)] that can be taken to make precise the claim that the geodesic principle is a theorem, rather than a postulate, of General Relativity.  In this brief note, I show, by explicit counterexample, that not only is a weaker energy condition than the one Geroch and Jang state insufficient to prove the theorem, but in fact a condition still stronger than the one that they assume is necessary.\\
\rule{6.5in}{1pt}\\[1\baselineskip]

\doublespacing

The status of the geodesic principle in General Relativity (GR), which states that free massive test point particles traverse timelike geodesics, has received considerable attention in the recent literature on the conceptual and mathematical foundations of spacetime theories.\footnote{See, for instance, \citet{Brown}, \citet{MalamentGeodesic}, \citet{WeatherallJMP}, and \citet{Sus}.}  This interest was prompted in large part by Harvey Brown's discussion of inertial motion in \emph{Physical Relativity} \citep{Brown}.  Much of the discussion has focused on a theorem originally due to Bob Geroch and Pong Soo Jang \citep{Geroch+Jang} that makes precise the claim that the geodesic principle can be understood as a theorem, rather than a postulate, of GR.  Following \citet[Prop. 2.5.2]{MalamentGR}, the Geroch-Jang result can be stated as follows.
\begin{thm}\label{GJ}\singlespacing
\emph{(\textbf{\citet{Geroch+Jang}})}
Let $(M,g_{ab})$ be a relativistic spacetime, with $M$ orientable.  Let $\gamma:I\rightarrow M$ be a smooth, imbedded curve.  Suppose that given any open subset $O$ of $M$ containing $\gamma[I]$, there exists a smooth symmetric field $T^{ab}$ with the following properties.
\begin{enumerate}
\item \label{sdec} $T^{ab}$ satisfies the \emph{strengthened dominant energy condition}, i.e. given any timelike covector $\xi_a$ at any point in $M$, $T^{ab}\xi_a\xi_b\geq 0 $ and either $T^{ab}=\mathbf{0}$ or $T^{ab}\xi_a$ is timelike;
\item \label{cons}$T^{ab}$ satisfies the \emph{conservation condition}, i.e. $\nabla_a T^{ab}=\mathbf{0}$;
\item \label{inside}$\supp{T^{ab}}\subset O$; and
\item \label{non-vanishing}there is at least one point in $O$ at which $T^{ab}\neq \mathbf{0}$.
\end{enumerate}
Then $\gamma$ is a timelike curve that can be reparametrized as a geodesic.
\end{thm}

Of particular interest has been the role of the strengthened dominant energy condition, condition \ref{sdec}, in proving the theorem.  The reason this condition is of interest is that a strong general assumption regarding the nature of matter appears to be at odds with the claim, apparently supported by some commentators (c.f. \citet{Brown} and \citet{Sus}, but also older works, such as \citet{Carmeli}), that the geodesic principle is a consequence of (just) the geometrical and geometro-dynamical structure of GR (including Einstein's field equation).  The status of this energy condition was clarified by \citet{MalamentGeodesic}, who showed the following.
\begin{thm}\singlespacing\emph{\textbf{\citep[Prop. 2.5.3]{MalamentGR}}} Let $(M,g_{ab})$ be Minkowski spacetime, and let $\gamma:I\rightarrow M$ be \emph{any} smooth timelike curve.  Then given any open subset $O$ of $M$ containing $\gamma[I]$, there exists a smooth symmetric field $T^{ab}$ on $M$ that satisfies conditions
\ref{cons}, \ref{inside}, and \ref{non-vanishing} of Theorem \ref{GJ}.\end{thm}
Malament's result shows that the energy condition is necessary, in the sense that the other three conditions together are not sufficient to prove the theorem.  Indeed, at least in Minkowski space, matter satisfying conditions \ref{cons} - \ref{non-vanishing} can be constructed in arbitrarily small neighborhoods of any timelike curve at all.

In this short note, I offer a further remark on the status of this energy condition.  It is a small point, but I think it is nonetheless worth making, if only to lay out the terrain for future discussions on this topic.  The remark concerns a question that arises in conjunction with Malament's result.  While Malament shows that some additional condition on $T^{ab}$, besides conditions \ref{cons} - \ref{non-vanishing}, is necessary, he does not prove that the full strengthened dominant energy condition is necessary.  One might thus wonder whether a weaker energy condition would be sufficient.

Resolving this question is of some foundational interest.  The point is most striking in the context of recent work on the status of the geodesic principle in geometrized Newtonian gravitation \citep{WeatherallJMP}, since there, too, an energy condition of sorts is necessary.  That condition, sometimes called the mass condition, is the requirement that mass density $\rho=T^{ab}t_at_b$, where $t_a$ is the temporal metric of geometrized Newtonian gravitation, is strictly positive.  One might take this condition to be a benign and unsurprising characterization of what we mean by ``massive particle'' in Newtonian gravitation \citep{WeatherallGP}.  If one does so, it is natural to ask if the corresponding energy condition in the Geroch-Jang theorem supports a similar interpretation.  The Newtonian mass condition is most closely analogous, at least superficially, to the so-called ``weak energy condition'' in GR, which requires that for any timelike covector field $\xi^a$, $T^{ab}\xi_a\xi_b\geq 0$.  The weak energy condition, like the mass condition, is naturally interpreted as the requirement that energy-momentum as determined by any observer is always non-negative; it is to be contrasted with the strictly stronger strengthened dominant energy condition, which additionally requires that the four-momentum of a matter field, as determined by any observer, be timelike (no such additional constraint is required in the Newtonian case).  Given these considerations, it seems salient to ask whether the weak energy condition, in conjunction with conditions \ref{cons} - \ref{non-vanishing}, is sufficient to prove the Geroch-Jang theorem, since if so, one might similarly interpret the energy condition in the Geroch-Jang theorem as part of what we mean by a massive particle.\footnote{At least two people with whom I have discussed this topic have suggested that the interpretation that the energy condition captures what we mean by a massive particle is merited even if the strengthened dominant energy condition is required, since what we mean by a massive particle in GR is a particle with positive mass that propagates causally.  Perhaps this is right---but if so, the theorem in the Newtonian case is all the more intriguing, since in that context one gets causal propagation, in addition to geodesic motion, for free from the weaker condition.  One way or the other, one is left with the question of why matter propagates causally in GR in the first place.}

\begin{figure}\begin{center}
\includegraphics[width=.7\columnwidth]{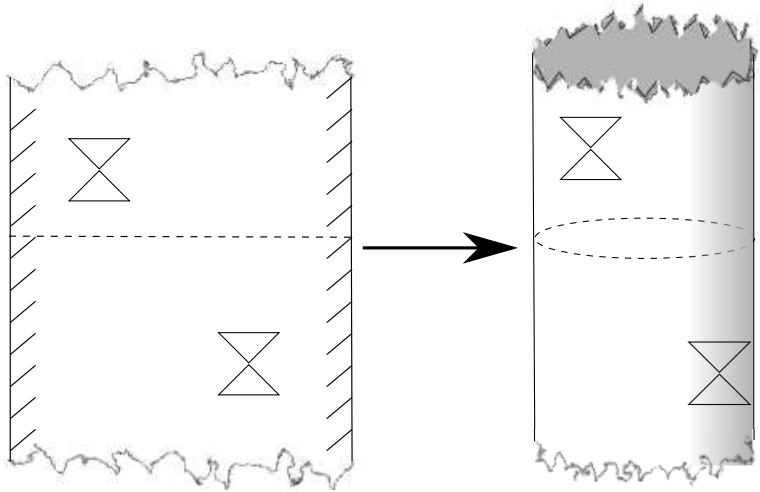}\end{center}
\caption{\label{fig1} The spacetime $(M,g_{ab})$ is constructed by taking 2 dimensional Minkowski spacetime and wrapping it up into a cylinder in the spacelike direction.  The metric remains the same, and so the lightcones are unchanged by the transformation.  The integral curves of $x^a$, however, are now closed curves.}
\end{figure}

The answer to this question is no, as can be seen from the following proposition.\footnote{I am indebted to David Malament for this formulation of the proposition.}
\begin{prop}\label{wec}\singlespacing There exist a relativistic spacetime $(M,g_{ab})$ and a smooth, imbedded spacelike curve $\gamma:I\rightarrow M$ satisfying the following condition.  Given any open neighborhood $O$ containing the image of the curve, there exists a smooth, symmetric rank 2 tensor field $T^{ab}$ satisfying conditions $\ref{cons}-\ref{non-vanishing}$ relative to $O$ and also satisfying the weak energy condition.\end{prop}
Proof. Let $(N,\eta_{ab})$ be 2 dimensional Minkowski spacetime, and let $t,x:N\rightarrow \mathbb{R}$ be a global coordinate system on $(N,\eta_{ab})$ relative to which the metric takes the form $\eta=\text{diag}(1,-1)$.\footnote{The following construction is easiest to picture in 2 dimensions, so I will develop it there. It should be clear however that the dimension does not actually play a role in what follows.}  We will take $M$ to be the manifold defined by associating points $(t,x)\in N$ and $(t,x+1)\in N$, yielding a cylinder (see Fig. \ref{fig1}).  We then take the metric $g_{ab}$ on $M$ to be pointwise equal to $\eta_{ab}$.  This will be the spacetime used to substantiate the existential claim made in the proposition.  Note that the $x$ and $t$ coordinates can be used to define two constant vector fields, which we will write as \begin{align*}
x^a=\left(\frac{\partial}{\partial x}\right)^a& &\text{and}& &t^a=\left(\frac{\partial}{\partial t}\right)^a.\end{align*}  Since $x^a$ is a constant spacelike field, its integral curves are spacelike geodesics; because of the topology of $M$, these are closed curves.  So pick some $o\in M$  (for convenience, suppose $t(o)=0$) and let $\gamma:I\rightarrow M$ be the maximal integral curve of $x^a$ through $o$.  This will be the curve described in the proposition.

It remains to show that with this choice of spacetime and curve, given any neighborhood $O$ of the curve, there exists a smooth, symmetric field $T^{ab}$ satisfying the four required conditions.  We will exhibit this field explicitly.  Let $O$ be any neighborhood of the curve.  Because the curve is closed, it must be possible find a minimal radius for $O$, that is, there must be some value $t_0$ such that the (closed) tube $\{p\in M : |t(p)| \leq t_0\}$ is a subset of $O$.  (This follows because the manifold is compact in the $x$ direction.)  Let  $\alpha$ be a smooth scalar field such that (a) $\alpha=1$ on $\gamma$, (b) $\alpha = 0$ outside of the (open) tube  $\{p\in M : |t(p)| < t_0\}$, (c) $\alpha \geq 0$ everywhere, and (d) $x^a\nabla_a\alpha=\mathbf{0}$.  One candidate for $\alpha$ would be the scalar field defined by \begin{equation}\label{alph}
\alpha(t)=\left(\frac{f(t+t_0)}{f(t+t_0)+f(-t-t_0/2)}\right)
 \times \left(\frac{f(t_0-t)}{f(t_0-t)+f(t-t_0/2)}\right),\end{equation}
where \[f(s)=\begin{cases} \exp(-1/s) &\text{if }s>0\\ 0 &\text{if }s\leq 0\end{cases}.\]
We can then define $T^{ab}=\alpha x^a x^b$.  $T^{ab}$ clearly satisfies conditions \ref{inside} and \ref{non-vanishing} by the construction of $\alpha$.  This $T^{ab}$ also satisfies the conservation condition, since $\nabla_a T^{ab}=x^bx^a\nabla_a(\alpha)=\mathbf{0}$, again by construction.  Finally, $T^{ab}$ also satisfies the weak energy condition, since for any timelike vector $\xi^a$, $T^{ab}\xi_a\xi_b=\alpha(x^a\xi_a)^2\geq 0$, since $\alpha\geq 0$.  Note, however, that it manifestly does not satisfy the strengthened dominant energy condition.\hspace{.25in}$\square$

The preceding proposition settles that the weak energy condition is not sufficient for the Geroch-Jang theorem.  But one can say even more.  Let me first draw attention to a subtle distinction between the energy condition stated above as condition \ref{sdec} of the Geroch-Jang theorem (what I call the ``strengthened dominant energy condition,'' following Malament) and the condition stated in the original Geroch-Jang paper, sometimes called the ``strict dominant energy condition''.  They are different.
\begin{quote}\singlespacing
\textbf{Strengthened Dominant Energy Condition}: An energy momentum field $T^{ab}$ satisfies the strengthened dominant energy condition if, given any timelike covector $\xi_a$ at any point in $M$, $T^{ab}\xi_a\xi_b\geq 0 $ and either $T^{ab}=\mathbf{0}$ or $T^{ab}\xi_a$ is timelike.
\end{quote}
\begin{quote}\singlespacing
\textbf{Strict Dominant Energy Condition}: An energy momentum field $T^{ab}$ satisfies the strict dominant energy condition if, given any two co-oriented timelike covectors $\xi_a$ and $\eta_a$ at any point in $M$, either $T_{ab}=\mathbf{0}$ or $T^{ab}\xi_a\eta_b>0$.
\end{quote}
The strengthened dominant energy condition is strictly stronger, as can be seen from the following equivalent formulation:
\begin{quote}\singlespacing
\textbf{Strengthened Dominant Energy Condition$^*$}: An energy momentum field $T^{ab}$ satisfies the strengthened dominant energy condition if, given any two co-oriented causal covectors $\xi_a$ and $\eta_a$ at any point in $M$, either $T_{ab}=\mathbf{0}$ or $T^{ab}\xi_a\eta_b>0$.
\end{quote}
With this reformulation, the strict dominant energy condition is a restriction on the product of $T^{ab}$ with pairs of co-oriented \emph{timelike} vector fields; the strengthened dominant energy condition is a restriction on the product of $T^{ab}$ with pairs of the larger class of future-directed \emph{causal} vector fields, which is more restrictive.

\begin{figure}\begin{center}
\includegraphics[width=.7\columnwidth]{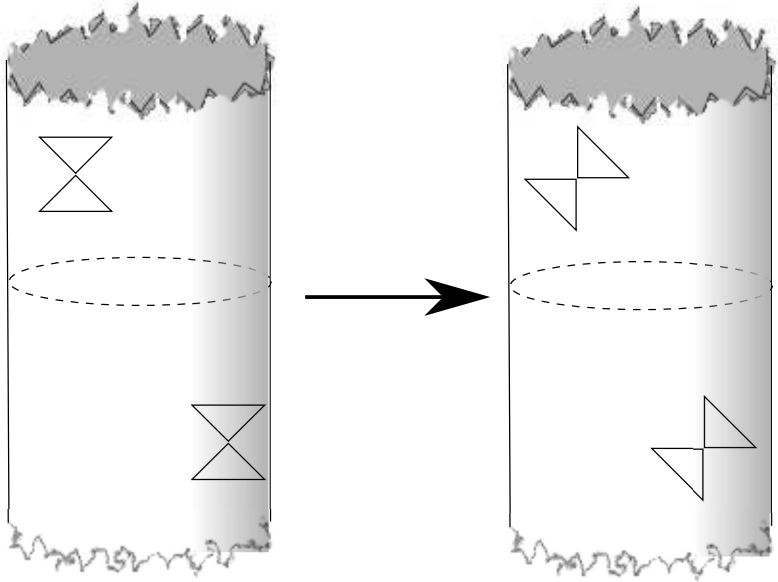}\end{center}
\caption{\label{fig2} The spacetime $(M,g'_{ab})$ is constructed by taking $(M,g_{ab})$, the cylindrical spacetime described in proposition \ref{wec}, and rotating the lightcones so that $x^a$ and $t^a$ are null fields with respect to the new metric.  The integral curves of $x^a$, however, remain closed curves.}
\end{figure}

It turns out that the strict dominant energy condition, already a strong energy condition by any standard, is still not strong enough.
\begin{prop}\singlespacing There exists a relativistic spacetime $(M,g_{ab})$ and a smooth, imbedded null curve $\gamma:I\rightarrow M$ satisfying the following condition.  Given any open neighborhood $O$ containing the image of the curve, there exists a smooth, symmetric rank 2 tensor field $T^{ab}$ satisfying conditions $\ref{cons}-\ref{non-vanishing}$ relative to $O$ and also satisfying the strict dominant energy condition.\end{prop}
Proof.  We will begin with   the cylindrical spacetime $(M,g_{ab})$ and coordinate system $t,x$ defined in the proof of the last proposition, modified as follows: we now consider a new metric, $g'_{ab}$, defined by: \[g'_{ab}=\frac{1}{2}\left((d_a t)(d_b x)+(d_b t)(d_a x)\right).\]  This new spacetime, $(M,g'_{ab})$, can be thought of as the original spacetime with the lightcones rotated so that, with respect to $g'_{ab}$, the constant vector fields $t^a$ and $x^a$ are now \emph{null} vector fields (see Fig. \ref{fig2}).  Relative to the new metric, the curve $\gamma$ we considered in proposition \ref{wec} is now a null curve (because its tangent vector field is now everywhere null).  From here we proceed identically to in proposition \ref{wec}.  Once again, in any neighborhood of $\gamma$, we can construct an energy-momentum field $T^{ab}=\alpha x^a x^b$, with $\alpha$ defined as in Eq. \eqref{alph}.  Now, however, $x^a$ is null, and so the field satisfies the strict dominant energy condition.\hspace{.25in}$\square$

These results indicate that one requires a strong energy condition indeed to prove the geodesic principle as a theorem of GR, at least by the Geroch-Jang method.  It is perhaps interesting to point out that the two propositions proved here differ in an important way from Malament's result: Malament shows that if one drops the energy condition altogether, one does not even get geodesic motion, much less timelike geodesic motion.  My focus has been slightly different, on how strong an energy condition is required to prove that the propagation of test matter is timelike (as opposed to spacelike or null).  In both cases, my counter-examples have involved geodesics (either spacelike geodesics, or null geodesics).  I do not know whether some weaker energy condition, say the weak energy condition, is sufficient to prove geodesic motion for test particles (without prejudice for whether the worldlines are timelike, spacelike, or null).  It may be that this would be an interesting question to settle.  That said, at least for some foundational purposes, it seems to me that one is most interested in the conditions that are necessary to prove the geodesic principle as ordinarily understood (i.e., as concerning timelike curves).

\singlespacing

\end{document}